\documentclass[twocolumn,prd,amsmath,amssymb,superscriptaddress,showpacs,nofootinbib]{revtex4}
\usepackage[T1]{fontenc}
\usepackage[latin9]{inputenc}
\usepackage{graphicx}
\usepackage{amssymb}
\usepackage{esint}

\makeatletter
\@ifundefined{textcolor}{}
{%
 \definecolor{BLACK}{gray}{0}
 \definecolor{WHITE}{gray}{1}
 \definecolor{RED}{rgb}{1,0,0}
 \definecolor{GREEN}{rgb}{0,1,0}
 \definecolor{BLUE}{rgb}{0,0,1}
 \definecolor{CYAN}{cmyk}{1,0,0,0}
 \definecolor{MAGENTA}{cmyk}{0,1,0,0}
 \definecolor{YELLOW}{cmyk}{0,0,1,0}
 }

\makeatother

\begin{document}

\title{Magnetic Tuning of the Relativistic BCS-BEC Crossover}

\author{Jin-cheng Wang}

\affiliation{Interdisciplinary Center for Theoretical Study and Department of
Modern Physics, University of Science and Technology of China, Hefei
230026, People's Republic of China}

\author{Vivian de la Incera}
\affiliation{Department of Physics, University of Texas at El Paso, El Paso, Texas
79968, USA}

\author{Efrain J. Ferrer}

\affiliation{Department of Physics, University of Texas at El Paso, El Paso, Texas
79968, USA}

\author{Qun Wang}

\affiliation{Interdisciplinary Center for Theoretical Study and
Department of Modern Physics, University of Science and Technology
of China, Hefei 230026, People's Republic of China}
\begin{abstract}
The effect of an applied magnetic field in the crossover from
Bose-Einstein condensate (BEC) to Bardeen-Cooper-Schrieffer (BCS)
pairing regimes is investigated. We use a model of relativistic
fermions and bosons inspired by those previously used in the context
of cold fermionic atoms and in the magnetic-color-flavor-locking
phase of color superconductivity. It turns out that as with cold atom
systems, an applied magnetic field can also tune
the BCS-BEC crossover in the relativistic case. We find that no matter what the initial state is at $B=0$, for large enough magnetic fields the system always settles into a pure BCS regime. In contrast to the atomic case, the
magnetic field tuning of the crossover in the relativistic system is
not connected to a Feshbach resonance, but to the relative numbers
of Landau levels with either BEC or BCS type of dispersion
relations that are occupied at each magnetic field strength.
\pacs{12.38.Mh, 03.75.Nt, 24.85.+p, 26.60.-c}

\end{abstract}
\maketitle

\section{Introduction}
In recent years many experimental advances have been made in pairings
of ultracold fermionic atoms, where the effective attractive interaction
between the atoms can be tuned with the help of an applied magnetic
field via a Feshbach resonance \cite{Exp1}. By tuning the fermion-fermion
interaction it has been possible to experimentally realize the crossover
between the weakly coupled Bardeen-Cooper-Schrieffer (BCS) superfluid
regime with the formation of Cooper pairs of two fermionic atoms and
the strong coupling regime where the pairs turn into difermion molecules
in Bose-Einstein condensation (BEC) \cite{Exp}. Even though there
is no phase transition but just a crossover between these two regimes,
their features are very distinct. In the BCS side the coherence length
of the pairs is much larger than the mean interparticle distance and
as a consequence the fermionic degrees of freedom are still manifested.
However, in the BEC side, the strong interaction allows two fermions
to bound into a bosonic molecule; thus no fermionic degrees of freedom
remain.

The BCS-BEC crossover is not limited to cold fermionic atoms or to
nonrelativistic systems. The main ingredients -- a dilute gas of
fermions with an attractive interaction that can favor the formation
of Cooper pairs on the Fermi surface and a viable mechanism to produce
the crossover -- can be found in a wide range of cold and dense fermion
systems. These conditions can be naturally satisfied inside the core
of neutron stars, where temperatures are relatively low compared to
densities which can reach values several times the normal nuclear
density and hence allow deconfinement. The conditions for a BCS-BEC
crossover can also be expected to be met in the planned low-energy
experiments at the Relativistic Heavy Ion Collider (RHIC) and future facilities all over the world,
such as the Facility for Antiproton and Ion Research (FAIR) \cite{Senger et al}, the Nuclotron-Based Ion Collider Fcility (NICA), or the Japn Proton Accelerator Research Complex (J-PARC) \cite{Fuku-Hatsuda2010}.

In recent years, interest has been spurred in investigating
the realization of the BCS-BEC crossover in various QCD-inspired
models \cite{Rel-BEC}-\cite{Basler:2010xy}. A strong motivation for
this activity is the need to explore the QCD-phase map at
intermediate densities and low temperatures, a region of significant
relevance for the physics of compact stars but inaccessible with
lattice QCD due to the complex fermion determinant. The phase of QCD
at asymptotically high baryonic densities is well established to be
a color superconducting (CS) phase \cite{csc}. This CS phase is the
result of the attractive color force in the antitriplet channel for
two quarks which favors the formation of Cooper pairs on the Fermi
surface. However, as the density decreases, the quark-gluon
interaction becomes stronger leading to a reduction of the coherence
length of the diquark pairs. What happens at this point is still a
matter of debate. Some model calculations
\cite{Asakawa'89}-\cite{Berges'99} suggest that the quark matter
might go directly to a chirally broken hadronic phase via a strong
first-order transition. Another possibility is that the diquark
pairs turn first into diquark molecules, thereby undergoing
 a BCS-BEC crossover \cite{Coherence-Length,Abuki:2010jq}. Eventually, the diquark
pair may pick up yet another quark to form a color-singlet baryon. Hence,
the BCS-BEC crossover could hold the key to our understanding of the
transition from CS to hadronic matter. In 1999 Schafer and Wilczek
\cite{Schafer-Wilczek 99} conjectured that the transition from CS
to hadronic matter should be actually a crossover. The quark-hadron
continuity has been studied in terms of the spectral continuity of
Nambu-Goldstone modes \cite{Yamamoto et al 07} and vector mesons
\cite{Hatsuda etal 08}. The role of diquarks in baryon formation
and dissociation in cold dense quark/nuclear matter has been recently
studied in Ref. \cite{WWR2010}.

Up to now, nevertheless, one ingredient has been left out in all the
investigations of the CS-hadronic matter transition via a BCS-BEC
crossover: an external magnetic field. However, magnetic fields are
endemic in neutron stars. Pulsars' magnetic fields range between
$10^{12}$ and $10^{13}$ G \cite{Pulsars}, and for magnetars they can
be as large as $10^{14}-10^{16}$ G \cite{Magnetars} on the surface
and presumably much larger in the core. Upper limit estimates for
neutron star magnetic fields indicate that their magnitude can reach
$\sim10^{18}-10^{20}$ G \cite{virial}-\cite{EoS-B2010}. Very strong
magnetic fields, $\sim10^{18}$ G/$10^{19}$ G, are also generated in
heavy-ion collisions at RHIC and LHC \cite{RHIC-B}. Nonetheless,
these experiments produce a hot and low-density matter that is far
from the QCD-phase region where the BCS-BEC crossover is expected to
occur. On the other hand, as already mentioned, the future
low-energy experiments at RHIC, NICA and FAIR \cite{Senger et
al}-\cite{Fuku-Hatsuda2010} have been designed to probe the phase
diagram of nuclear matter at intermediate-to-large baryon density and low
temperature. These experiments are expected to produce also very
strong magnetic fields \cite{future exp}, hence they will be
relevant for understanding the field's influence on the CS-hadronic
transition.

Because of the astrophysical relevance, and also in preparation for
those future experiments, it is important to have a good theoretical
understanding of the magnetic field effects on the CS-hadronic
matter crossover. The present paper is a first attempt in this
direction. It is remarkable that the identification between the low-energy theories of the hadronic matter and the color-flavor-locking (CFL) phase
\cite{Schafer-Wilczek 99}, which served as the base for the
quark-hadronic matter continuity conjecture, was later found to
exist too in an external magnetic field \cite{MCFL}. In this case
the identification was between the low-energy modes of the magnetic
CFL (MCFL) phase \cite{MCFL} and those of the hadronic matter in a
magnetic field \cite{miransky-shovkovy}. We hope that the results of
the present paper will shed some light on the quark-hadronic matter
crossover in the presence of a magnetic field.

The most important outcome of this work is the discovery of a new
mechanism by which a magnetic field can tune the BCS-BEC crossover.
The mechanism is related to the filling (emptying) of new Landau
levels (LLs) when the field is varied and to the relative numbers of
occupied LLs with either BEC or BCS type of dispersion relations at
a given magnetic field value. The filling (emptying) of new LLs with
varying field is also responsible for the de Haas-van Alphen
oscillations of the gap \cite{osc-gap}-\cite{oscillations} and
number densities. No matter what the initial state of the system is
at $B=0$, for large enough magnetic fields the system will always
reach a pure BCS regime.

Even though our calculation is based on a simple model, it
encompasses the properties of spin-zero CS that are essential for
the new tuning mechanism to work, mainly that the pairing fermions
carry opposite charges (equivalent to the rotated charge in CFL and
2SC) to ensure the coupling of these fermions with the external
field, and the lack of a Meissner effect. Moreover, the
field-induced tuning mechanism is model-independent. The crossover
to the BCS regime at strong field strengths occurs because at those
fields most of the fermions will lie in their lowest Landau level
(LLL) and the dispersion relation of the LLL quasiparticles in the
paired system is always of BCS type. Notice that this mechanism is
different from the Feshbach resonance that produces the crossover in
cold atom systems \cite{Interaction} by tuning the effective
interaction between the fermions.

The plan of the paper is the following. In Sec. II we introduce
the model and derive the gap and chemical equilibrium equations. In
Sec. III we present our numerical results and discuss their
meaning, as well as the physical origin of the crossover at large magnetic
fields. The concluding remarks are given in Sec. IV.

\section{Relativistic Fermion-Boson Model in a Magnetic Field}
To explore the effects of the magnetic field on the BCS-BEC crossover,
we will extend the model of fermions and scalar bosons interacting via
a Yukawa term considered in \cite{Q-Wang}, to allow for two
oppositely charged fermions $\Psi^{T}=(\psi_{1},\psi_{2})$ that couple
to an external, uniform and constant magnetic field B. The symmetry
group of the model is $\mathrm{U(1)_{B}}\otimes\mathrm{U(1)_{em}}$, with subscripts
"B" and "em" labeling the groups of baryonic and electromagnetic transformations
respectively. The charged fermions in our model mimic the rotated
charged quarks that pair to form neutral Cooper pairs in the CFL and
2SC phases. The theory is described by the Lagrangian density \begin{equation}
{\cal L}={\cal L}_{f}+{\cal L}_{b}+{\cal L}_{I},\label{Lagrangian}\end{equation}
 with \begin{subequations} \begin{equation}
{\cal L}_{f}=\overline{\Psi}(i\gamma^{\mu}\partial_{\mu}+\mu\gamma^{0}-\widehat{Q}\gamma^{\mu}A_{\mu}-m)\Psi,\end{equation}
 \begin{equation}
{\cal L}_{b}=(\partial_{\mu}+2i\mu\delta_{\mu0})\varphi^{\ast}(\partial^{\mu}-2i\mu\delta^{\mu0})\varphi-m_{b}^{2}\varphi\varphi^{\ast},\end{equation}
 \begin{equation}
{\cal L}_{I}=\varphi\overline{\Psi}_{C}(i\gamma_{5}\widehat{G})\Psi+\varphi^{\ast}\overline{\Psi}(i\gamma_{5}\widehat{G})\Psi_{C}.\end{equation}
 \label{eq2} \end{subequations}
 Here $m$ and $m_{b}$ denote the
fermion and boson masses respectively. The charge-conjugate fermions
are described by $\Psi_{C}=C\overline{\Psi}^{T}$ with $C=i\gamma^{2}\gamma^{0}$,
and the electric charge $\widehat{Q}=q\sigma_{3}$ and Yukawa coupling
$\widehat{G}=g\sigma_{2}$ operators are given in terms of the Pauli
matrices $\sigma_{i}$. $A_{\mu}$ is the vector potential associated
with the external magnetic field \textit{B}, which, without loss of generality,
can be chosen along the $x_{3}$ axis.

The Lagrangian (\ref{Lagrangian}) is invariant under the
$\mathrm{U(1)_{B}}$ transformation
$\Psi\rightarrow\Psi'=e^{-i\alpha}\Psi$,
$\varphi\rightarrow\varphi'=e^{i2\alpha}\varphi$. Hence the bosons
carry twice the baryon number of the fermions. As in Ref.
\cite{Q-Wang}, chemical equilibrium with respect to the conversion
of two fermions into one boson and vice versa is ensured by
introducing fermion chemical potentials $\mu$ for fermions and
$2\mu$ for bosons. The transition can then be described from a
weakly coupled and neutral Cooper pair of two fermions with opposite
electric charges into a molecular difermionic bound state, with an
electrically neutral, strongly coupled, bosonlike behavior. In
order to describe the BEC of these molecules, we have to separate
the zero mode of the boson field $\varphi$ and replace it by its
expectation value $\phi\equiv\langle\varphi\rangle$, which
represents the electrically neutral difermion condensate. The
mean-field effective action is then \begin{eqnarray}
I^{B}(\overline{\psi},\psi) & = & \frac{1}{2}\int d^{4}x\, d^{4}y\,\overline{\Psi}_{\pm}(x){\cal S}_{(\pm)}^{-1}(x,y)\Psi_{\pm}(y)+\nonumber \\
 &  & +(4\mu^{2}-m_{b}^{2})\mid\phi\mid^{2}+\mid(\partial_{t}-2i\mu)\varphi\mid^{2}\nonumber \\
 &  & -\mid\nabla\varphi\mid^{2}-m_{b}^{2}\mid\varphi\mid^{2},\label{b-action}\end{eqnarray}
 where the fermion inverse propagators of the Nambu-Gorkov positive
and negative charged fields $\Psi_{+}=(\psi_{2},\psi_{1C})^{T}$ and
$\Psi_{-}=(\psi_{1},\psi_{2C})^{T}$ are given by \begin{equation}
{\cal S}_{(\pm)}^{-1}=\left(\begin{array}{cc}
[G_{(\pm)0}^{+}]^{-1} & i\gamma^{5}\Delta^{*}\\
i\gamma^{5}\Delta & [G_{(\pm)0}^{-}]^{-1}\end{array}\right)\ ,\label{inv-propg}\end{equation}
 with \begin{equation}
[G_{(\pm)0}^{\pm}]^{-1}(x,y)=[i\gamma^{\mu}\Pi_{\mu}^{(\pm)}-m\pm\mu\gamma^{0}]\delta^{4}(x-y)\ ,\label{B-x-inv-prop}\end{equation}
 and $\Pi_{\mu}^{(\pm)}=i\partial_{\mu}\pm qA_{\mu}$. We take the
external vector potential in the Landau gauge $A_{2}=Bx_{1}$, $A_{0}=A_{1}=A_{3}=0$.
The Bose condensate $\phi$ is related to the difermion
condensate through $\Delta=2g\phi$.

The zero temperature effective potential obtained from (\ref{b-action}) becomes, after using Ritus's transformation to momentum space \cite{Ritus},

 \begin{eqnarray}
\Omega=-\frac{qB}{2\pi^{2}}\sum_{e=\pm1}\sum_{k=0}^{\infty}d(k)\int_{0}^{\infty}dp_{3}\epsilon_{e}\qquad\qquad\nonumber \\
+\frac{(m_{b}^{2}-4\mu^{2})\Delta^{2}}{4g^{2}}+\frac{1}{4\pi^{2}}\sum_{e=\pm1}\int_{0}^{\infty}\omega_{e}p^{2}\, dp,\label{Eff-Pot}\end{eqnarray}
where $d(k)=(1-\frac{\delta_{k0}}{2})$ denotes the spin degeneracy of the Landau levels. The energy dispersions for fermions and bosons are
 \begin{subequations} \begin{equation}
\epsilon_{e}(k)=\sqrt{(\epsilon_{k}-e\mu)^{2}+\Delta^{2}},\qquad e=\pm1\quad\end{equation}
and
\begin{equation}
\omega_{e}=\sqrt{p^{2}+m_{b}^{2}}-2e\mu,\qquad e=\pm1\label{spectrum-b}\end{equation}
respectively. Index $k$ denotes the Landau level, $e$ labels quasiparticle/antiquasiparticle
contributions, and
 \begin{equation}
\epsilon_{k}=\sqrt{p_{3}^{2}+2|qB|k+m^{2}},\quad k=0,1,2,...
\end{equation}
\label{spectrum-f} \end{subequations}
is the energy of a free fermion in a magnetic field. As always occurs with charged fermions in a magnetic field, the transverse component of the momentum is quantized (Landau levels), so the 4-momentum of the fermion in the chosen gauge is given by $\overline{p}\equiv (p_{0},0,-sgn(qB)\sqrt{|2qB|k},p_{3})$ and the energy depends only on the longitudinal component $p_{3}$ (for a field parallel to $x_{3}$) and the Landau level $k$.

To investigate the crossover we first need to find the gap and chemical
potential that simultaneously solve the gap equation and the condition
of chemical equilibrium at fixed parameters, and then use them to
obtain the density fractions of fermions and bosons as functions of
the field.

Chemical equilibrium requires
\begin{equation}
n=n_{F}+n_{0}
\label{numberequa}\end{equation}
where $n$ plays the role of a fixed total baryon number density, $n=-\partial\Omega/\partial\mu$,
and the fermion number density $n_{F}$ and condensate density $n_{0}$,
are respectively given by
\begin{eqnarray}
n_{F} & = & -\frac{qB}{4\pi^{2}}\sum_{e=\pm1}\sum_{k=0}^{\infty}ed(k)\int_{0}^{\infty}dp_{3}\frac{\epsilon_{k}-e\mu}{\epsilon_{e}},\label{Fermion-density}\\
n_{0} & = & \frac{2\mu\Delta^{2}}{g^{2}}.\label{Condensate-density}
\end{eqnarray}

The gap equation is given by $\partial\Omega/\partial\Delta=0$,
which can be obtained from (\ref{Eff-Pot}) as
\begin{eqnarray}\label{Gap-Eq}
\frac{\widetilde{m}_b^2-4\mu^2}{2g^2}&=&
\frac{qB}{2\pi^2}\sum_{e=\pm 1}\sum_{k=0}^\infty d(k)\int_0^\infty
dp_3 \frac{1}{\epsilon_{e}(k)} \nonumber \\
&&-2\int \frac{d^{3}p}{(2\pi)^{3}} \frac{1}{\sqrt{p^{2}+m^{2}}}.
\end{eqnarray}
As discussed in \cite{Q-Wang}, the crossover parameter in the
present case can be defined by
$x\equiv-\frac{\widetilde{m}_{b}^{2}-4\mu^{2}}{2g^{2}}$, which is
linked to the renormalized boson mass $\widetilde{m}_{b}$ in vacuum
\begin{eqnarray}\label{Boson-mass}
\widetilde{m}_b^2=m_b^2-4g^2\int \frac{d^{3}p}{(2\pi)^{3}}
\frac{1}{\sqrt{p^{2}+m^{2}}}.
\end{eqnarray}
The parameter $x$ can then be changed by hand to mimic the
effect of a change in the coupling.

Since the momentum integral and the summation over fermion Landau
levels are divergent, we introduce a Gaussian regulator
$\exp[-(p_{3}^{2}+2|qB|k)/\Lambda^{2}]$ with momentum cutoff
$\Lambda=1$ GeV. Following the derivations of \cite{Q-Wang}, one can
see that at zero magnetic field the parameters of the theory $g$,
$n$, $m$, and $\widetilde{m}_{b}$ can be always chosen to have $x=0$
coinciding with the situation where the density fractions of
fermions $\rho_{F}=n_{F}/n$, and bosons $\rho_{b0}=n_{0}/n$ are all
equal to $1/2$. With such a choice, and according to the criterion
used in \cite{Q-Wang}, negative values of $x$ with large moduli
describe a pure BCS state, large positive values of $x$ describe a
pure BEC phase, and $1/x$ plays the role of the scattering length.
The selection of the model parameters can be done at any given
magnetic field value, to have $x=0$ corresponding to the unitarity
limit, at which the scattering length becomes infinite. In this
work, however, we are more interested in exploring the situation
where we keep fixed values of the parameters, and instead change the
strength of the magnetic field to see if it can have any effect in
the BCS-BEC crossover. Henceforth we will use $m=0.2$ GeV and $g=1$
in all the calculations \footnote{Although we are using a toy model
as a first attempt to study the field effects on the crossover, the
parameters of our model are consistent with the region of low
temperatures and intermediate densities, where the coupling is
expected to be relatively strong and chiral symmetry breaking can in
principle coexist with color superconductivity \cite{strong 2SC} .}.

\section{Results and Discussion }

\begin{figure}
\begin{centering}
\includegraphics[scale=0.53]{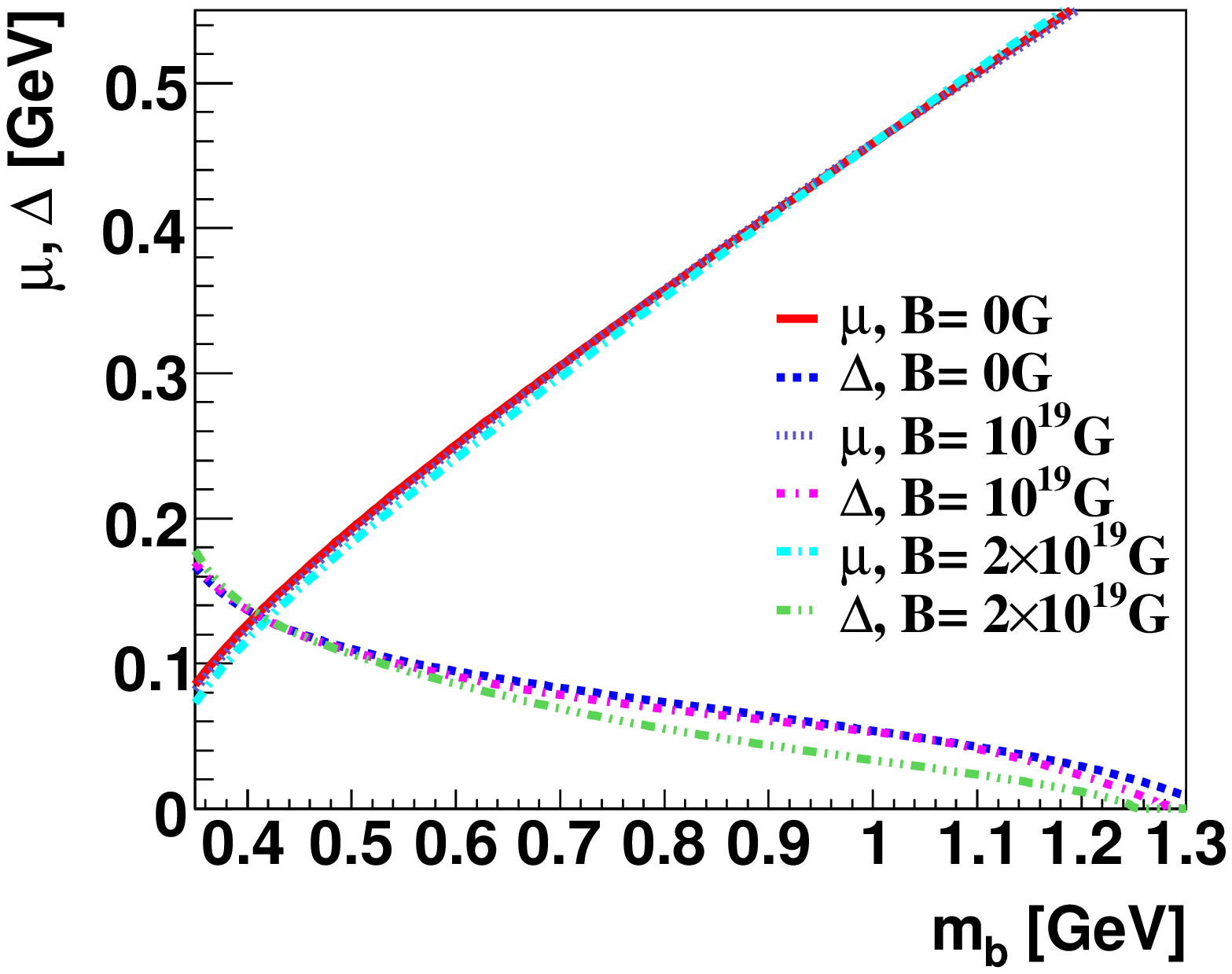}
\par\end{centering}

\begin{centering}
\includegraphics[scale=0.53]{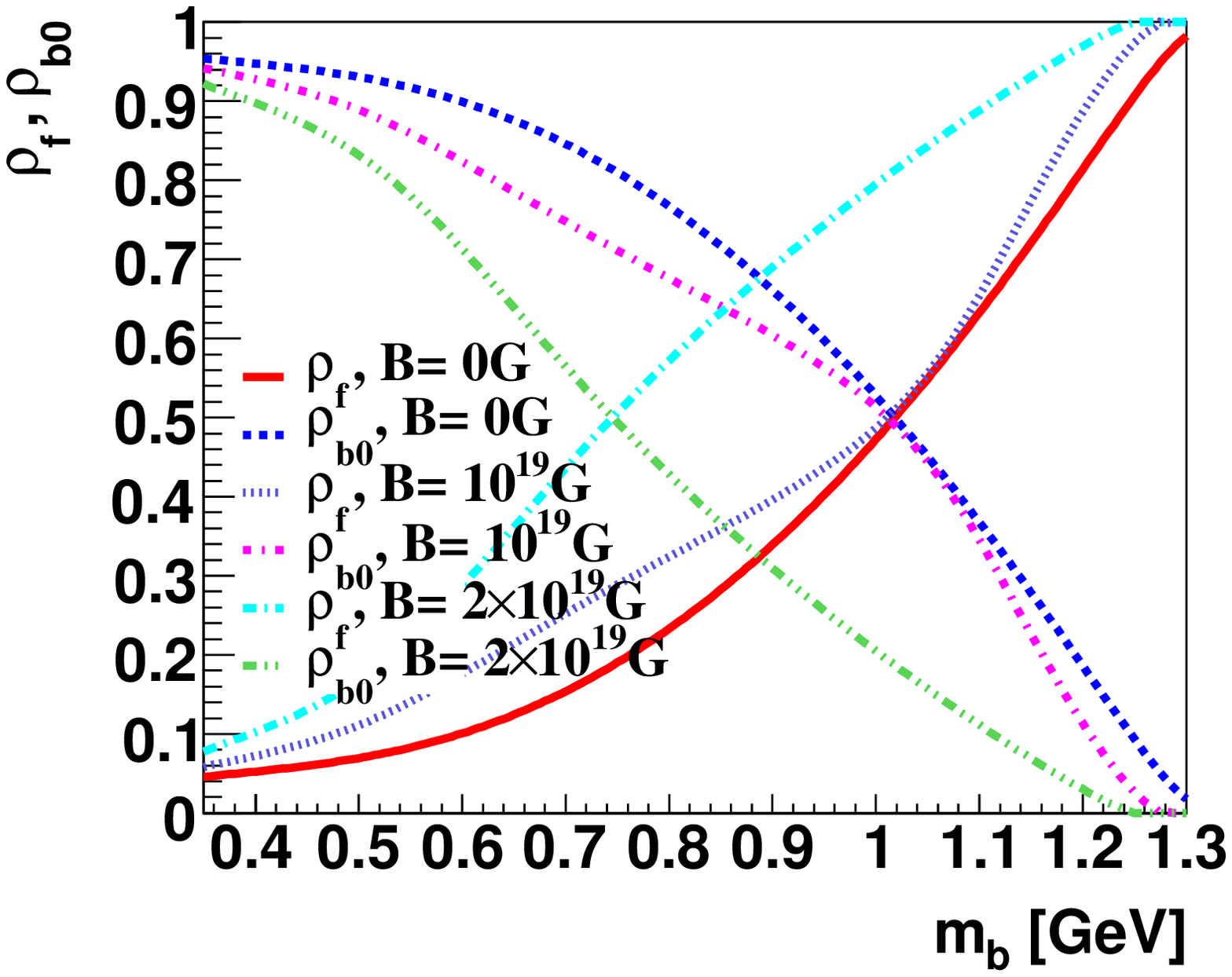}
\par\end{centering}

\centering{}\includegraphics[scale=0.53]{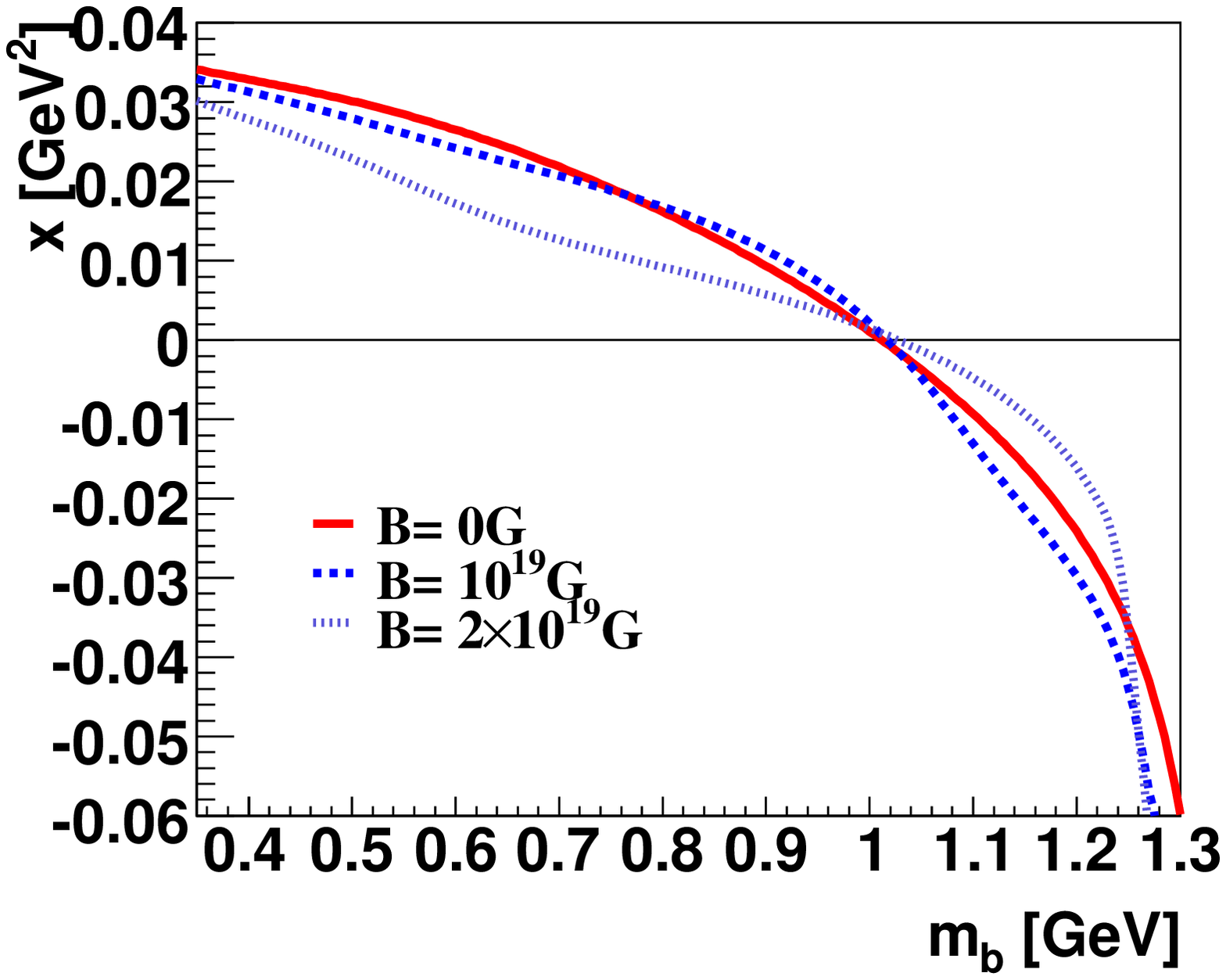}

\caption{{\footnotesize Various quantities as functions of $m_{b}$
at three values of the magnetic field, $B=0,10^{19},2\times10^{19}$
G: gap and fermionic chemical potential (upper panel), fermion and
boson number fractions (middle panel), the parameter $x$ (lower
panel). The cross point is
shifted by the magnetic field, as shown in the middle panel. The
system starts on the BEC side and crosses over to the BCS side when
$m_{b}$ increases. \label{fractionsversusmb} }}
\end{figure}

Let us discuss now the numerical solutions of Eqs.
(\ref{numberequa}) and (\ref{Gap-Eq}), along with the
field dependence of the physical quantities of the problem. In Fig.
\ref{fractionsversusmb}, we set the magnetic field to three values
$B=0,10^{19},2\times10^{19}$ G and study the influence of the field
on various quantities by tuning the renormalized boson mass
$\widetilde{m}_b$. Notice that for a fixed cutoff $\Lambda$,
changing the renormalized  boson mass, is equivalent to changing the
bare boson mass $m_{b}$.

\begin{figure}
\begin{centering}
\includegraphics[scale=0.45]{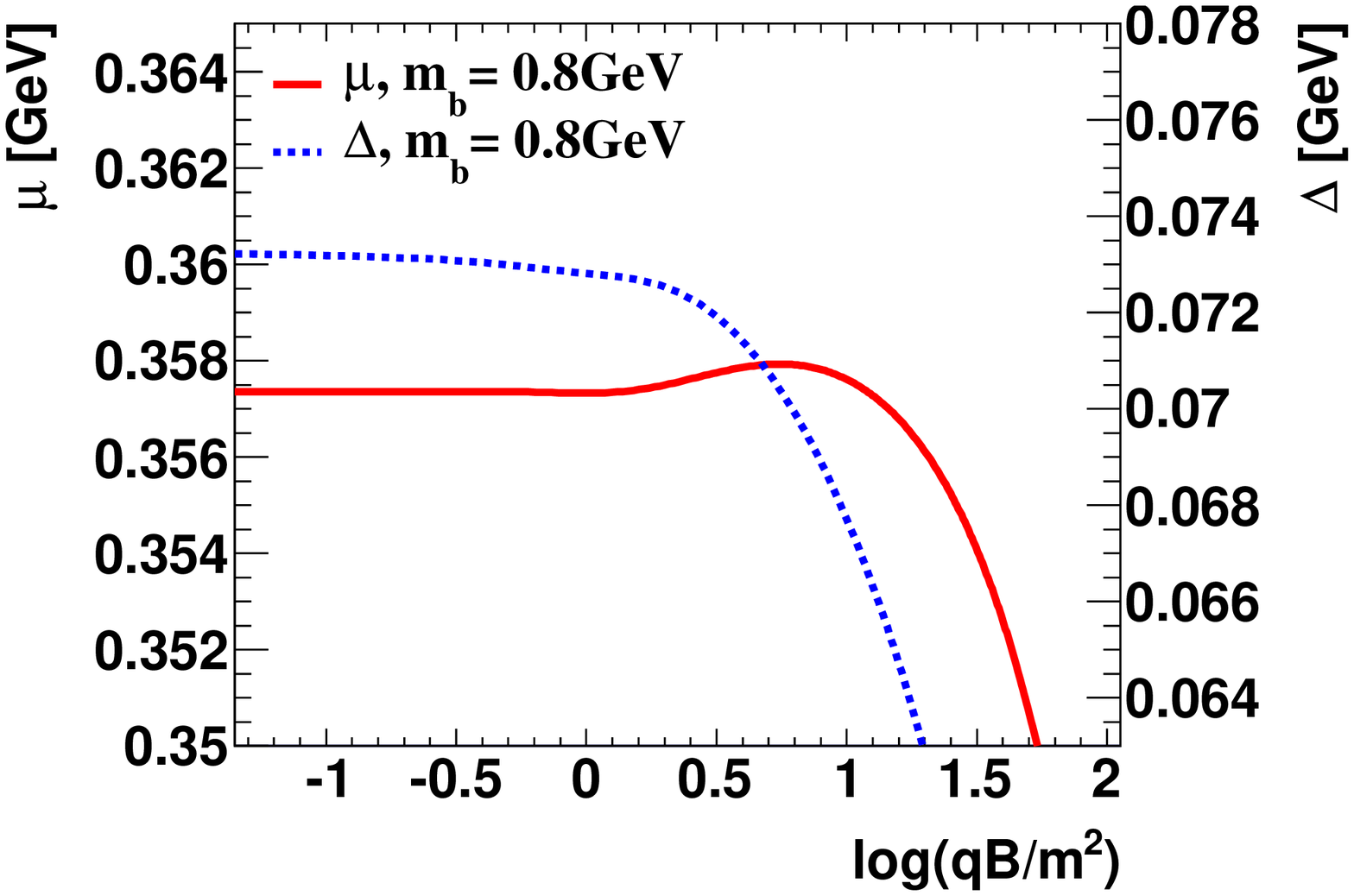}
\par\end{centering}

\begin{centering}
\includegraphics[scale=0.45]{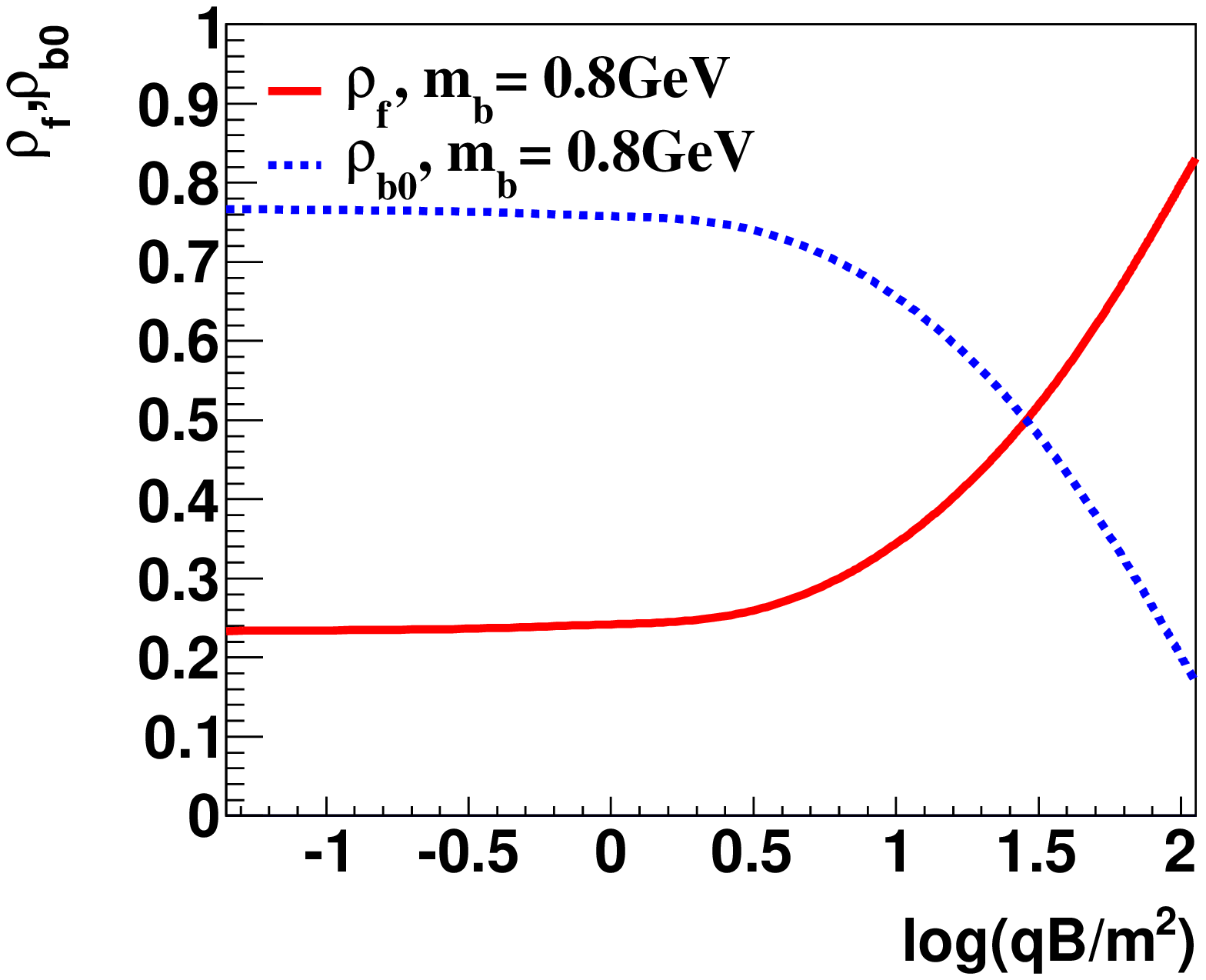}
\par\end{centering}

\centering{}\caption{{\footnotesize Various quantities as functions
of $\ln(|qB|/m^{2})$: gap and fermion chemical potential (upper
panel), fermion and boson number fractions (lower panel). Starting
from a BEC regime, on which the BEC component is much larger than the
BCS one at small fields on the left, the system
crosses over to a pure BCS state at large magnetic fields on the right. The scale of the field-induced oscillations is too small to be visible in the plot.
\label{fractionsversusB} }}

\end{figure}

\begin{figure}
\begin{centering}
\includegraphics[scale=0.45]{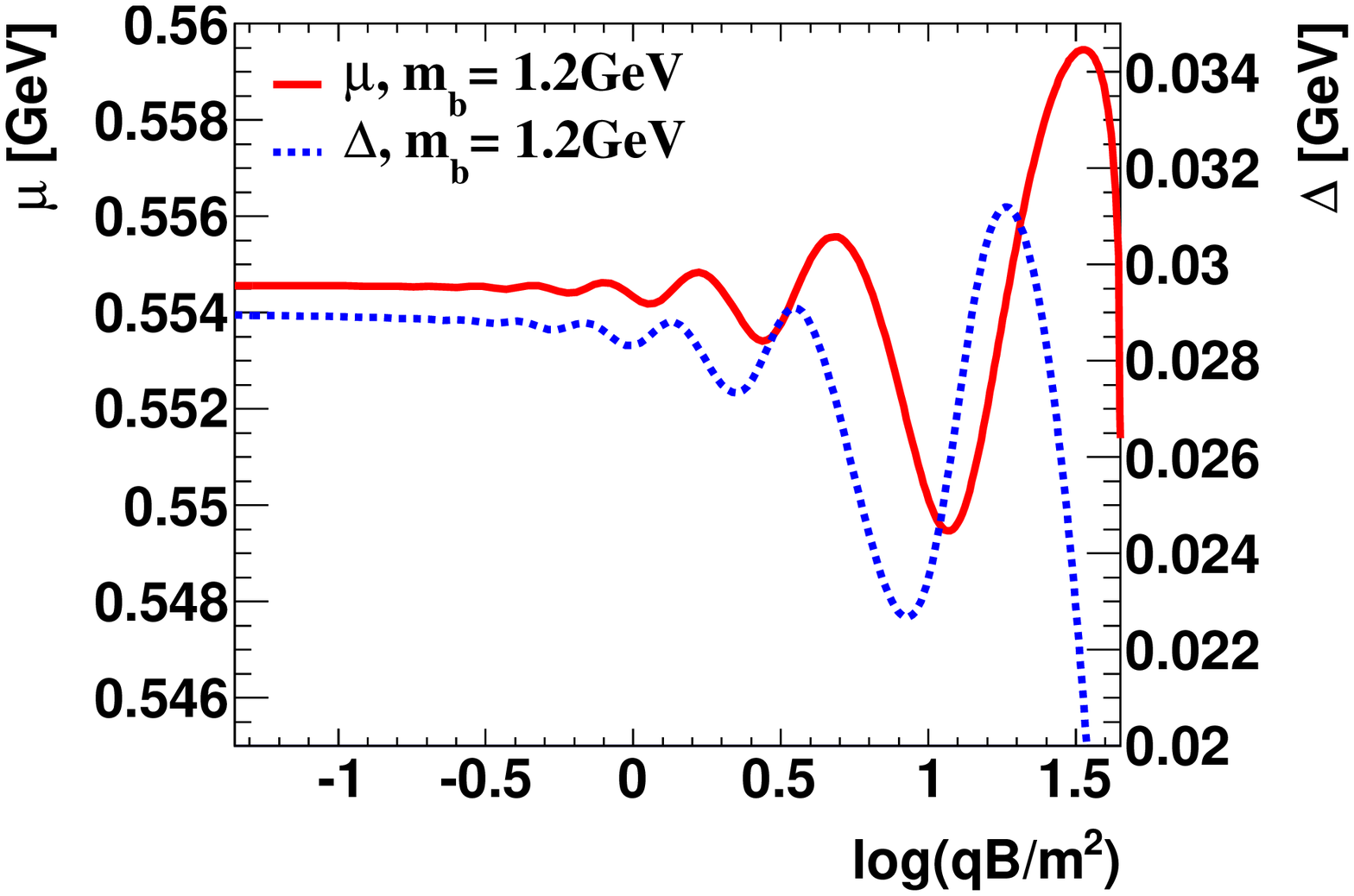}
\par\end{centering}

\begin{centering}
\includegraphics[scale=0.45]{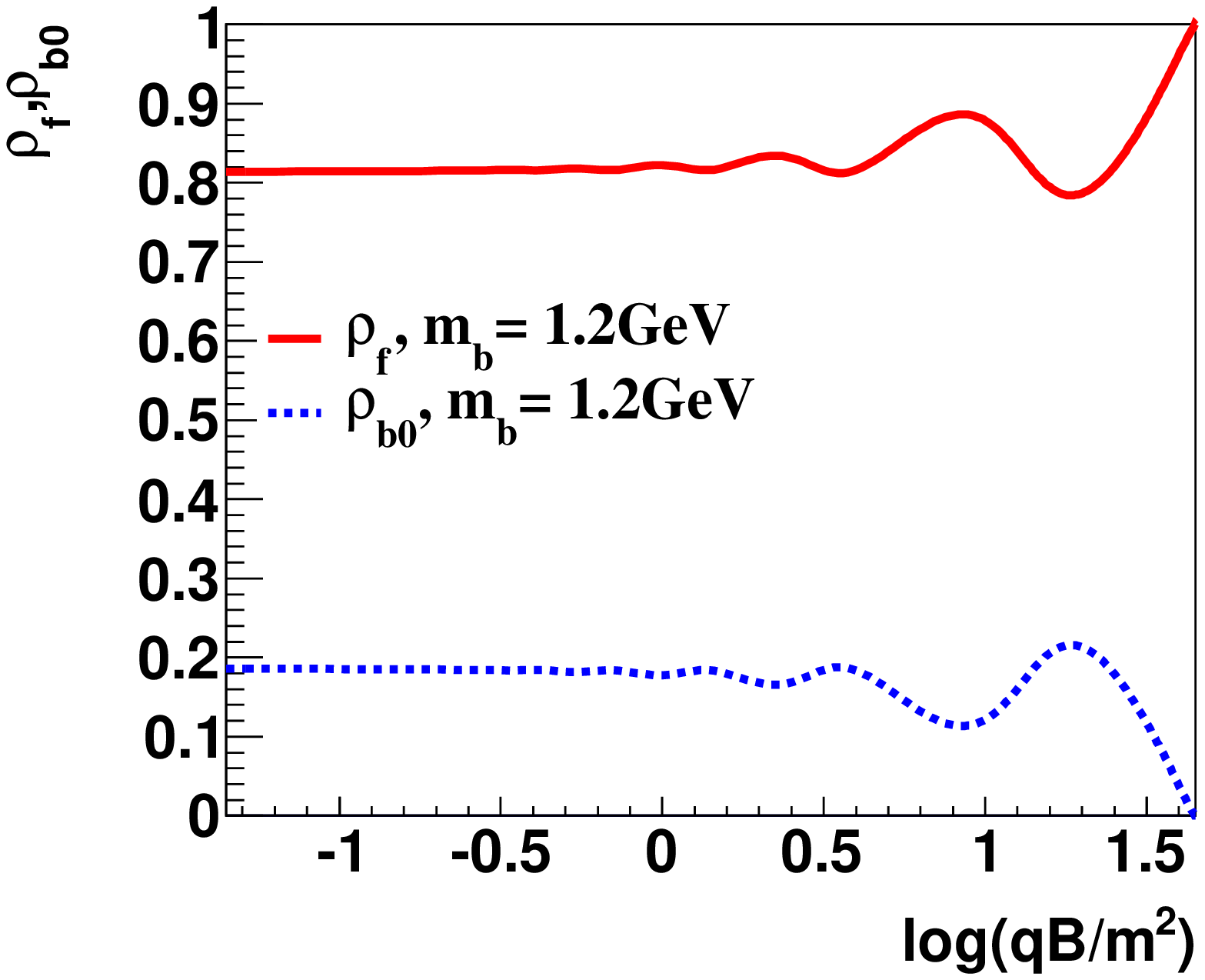}
\par\end{centering}

\centering{}

\caption{{\footnotesize Same as Fig. \ref{fractionsversusB} except
that the starting state has an excess of BCS pairing over BEC pairing at
small magnetic field. The magnetic field induces the de Haas-van Alphen oscillations in all the physical quantities with larger amplitudes than in
Fig. \ref{fractionsversusB} because the gap is smaller in this case. The oscillations stop when the field is large enough to put all the fermions in the LLL ensuring a pure BCS state.\label{fractionsversusB2} }}
\end{figure}

The upper panel in Fig.\ref{fractionsversusmb} shows the variations
of the chemical potential and gap solutions with the boson mass at
different field values. Even though at zero and nonzero magnetic
fields the chemical potential increases and the gap decreases with
$m_b$, a large magnetic field yields slightly smaller gap values for
the same $m_b$ in the large $m_b$ region. For the three field strengths,
the system is in a BEC state at $m_{b}\simeq 0.35$ GeV in the left
end region of the middle panel graph, where the BEC-like pairing
dominates over the BCS-like one, as reflected in the density
fractions, large $\rho_{b0}$ and small $\rho_{f}$. When $m_{b}$
increases, the fermion number fraction becomes larger and finally
dominates at the right end indicating a BCS regime. The system then
undergoes a crossover from BEC to BCS with increasing $m_b$. Notice
that the BCS-BEC crossover is realized in a similar way at zero and
nonzero magnetic fields. However, the external magnetic field tends
to favor BCS over BEC. When \textit{B} is large enough
($\gtrsim10^{19}$G),  the crossover point shifts to a lower $m_b$
and the system is in the BCS side for a much larger set of $m_{b}$
values. This strong magnetic field effect will become even more
apparent in the plots of Figs. \ref{fractionsversusB} and
\ref{fractionsversusB2}. The lowest panel in Fig.
\ref{fractionsversusmb} shows the variation of the parameter $x$
with $m_b$ at the three field values.

Figures \ref{fractionsversusB} and \ref{fractionsversusB2} present the
behavior of different parameters as functions of the magnetic field
when the system starts at zero field either in a BEC regime (Fig.
\ref{fractionsversusB}) or in a  BCS one (Fig.
\ref{fractionsversusB2}). The gap and the chemical potential solved from the gap
and density equations at zero magnetic field are $\Delta_{0}=0.073$ GeV
and $\mu_{0}=0.357$ GeV for $m_b=0.8$ GeV ($\widetilde{m}_b=0.692$ GeV), and $\Delta_{0}=0.029$ GeV and
$\mu_{0}=0.555$ GeV for $m_b=1.2$ GeV ($\widetilde{m}_b=1.131$ GeV). In both cases the system ends up in the
BCS regime at strong fields. As shown in Fig.
\ref{fractionsversusB}, the influence of the magnetic field on the
crossover is more dramatic when the system starts in the BEC side,
as reflected in the behavior of the number fractions shown in the lower panel of
that figure. On the other hand, when the system is in the BCS regime at $B=0$ (Fig. \ref{fractionsversusB2}) the applied magnetic field simply strengthens the nature of that regime. The de Haas-van Alphen oscillations displayed by the physical parameters in the two panels of Fig. \ref{fractionsversusB2} have been also obtained in other models of
color superconductivity in a magnetic field \cite{oscillations}. The oscillations also exist in Fig. \ref{fractionsversusB}, but their scale is smaller and hence are not visible in the plots. The scale of the oscillations is connected to the magnitude of the gap. The larger the $m_b$, the smaller the gap magnitude and hence, the larger the amplitude of the field-induced oscillations.

\begin{figure}

\begin{centering}
\includegraphics[scale=0.53]{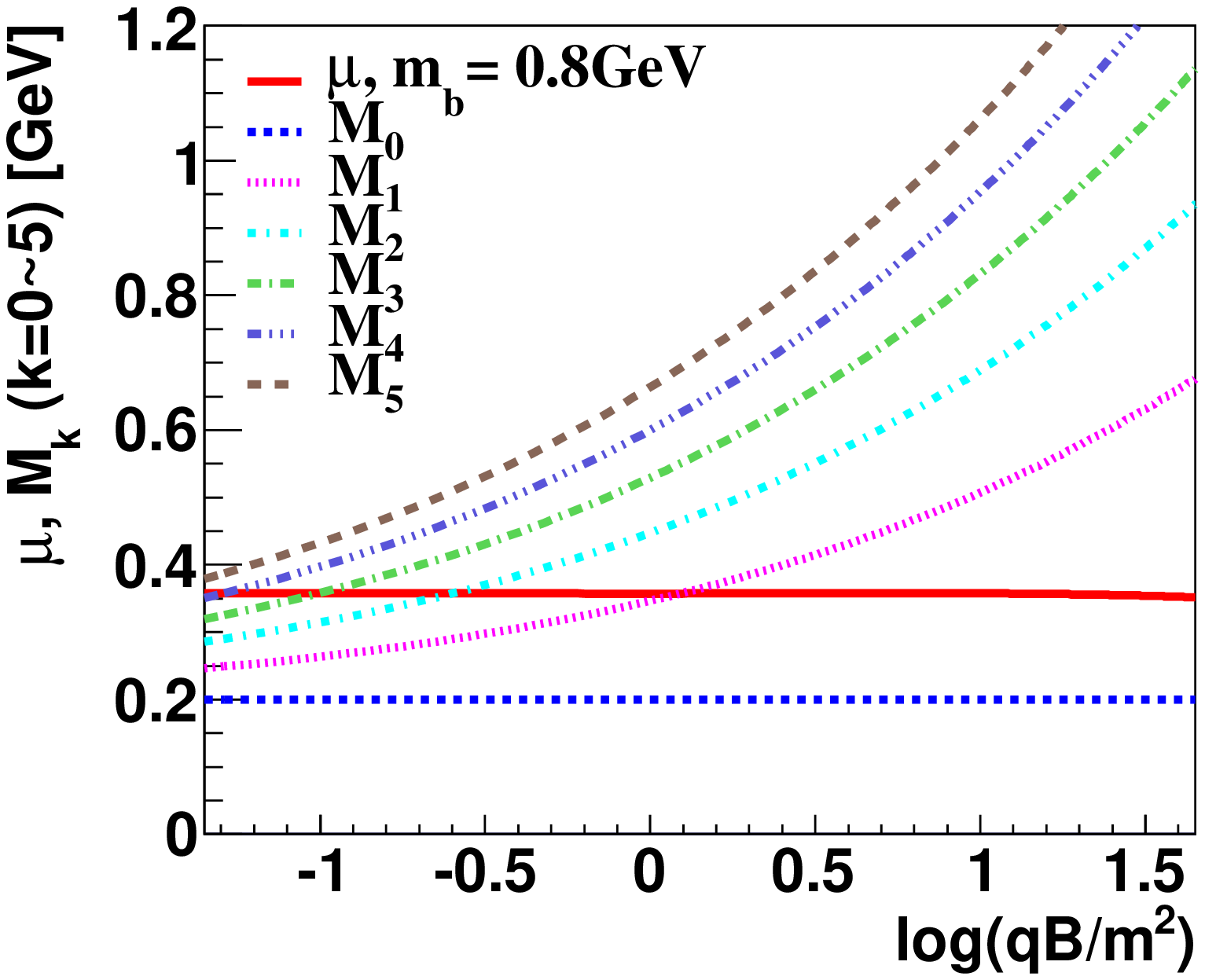}
\includegraphics[scale=0.53]{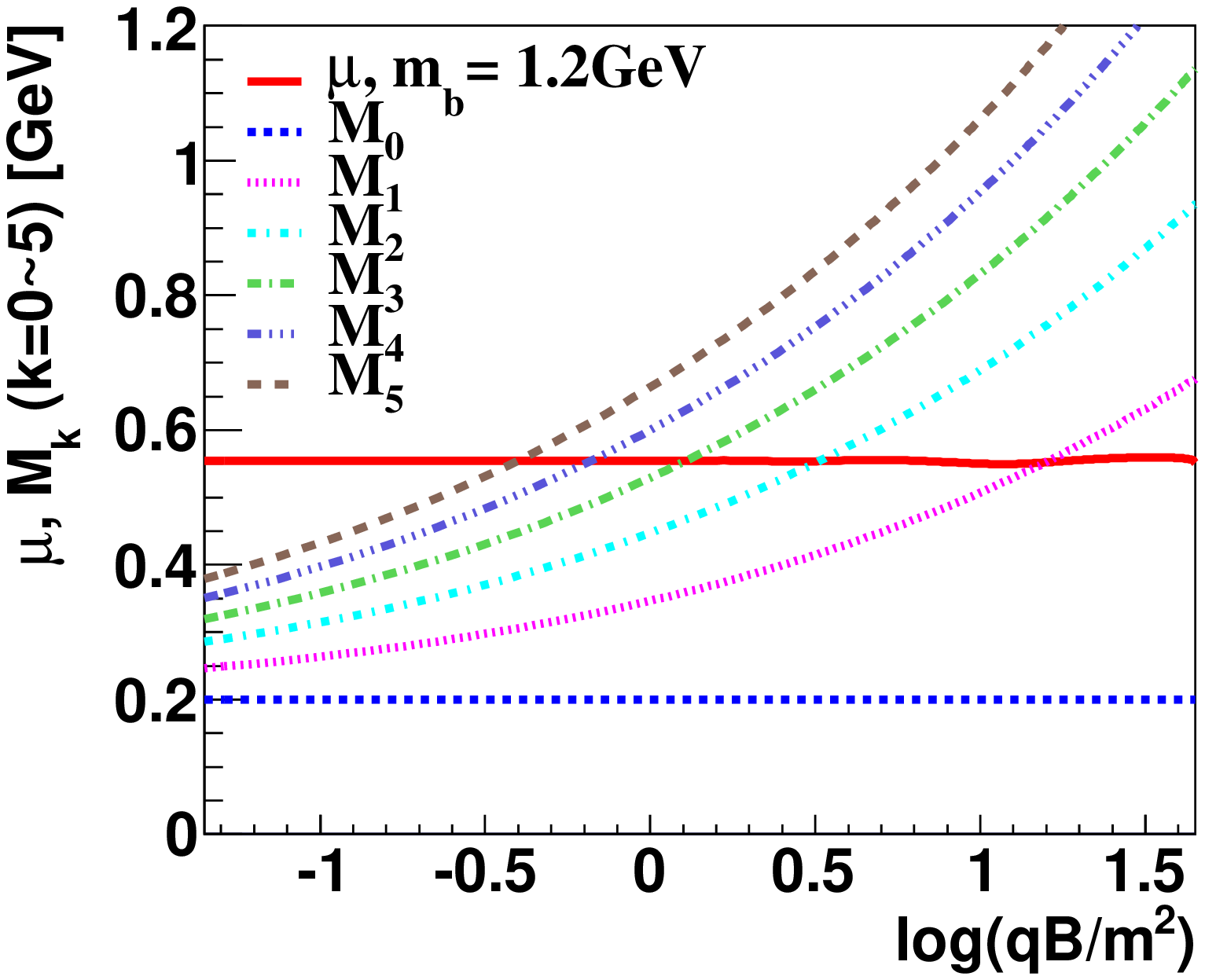}
\par\end{centering}

\caption{{Variation of the quasiparticles' effective masses of the first few Landau levels with the magnetic field. The solid (red) line represents the chemical potential. Masses over this line give rise to a bosonlike dispersion, those below the line to a fermionlike one. The upper (lower) panel corresponds to the case where the system starts in the BEC (BCS) regime at zero field. \label{effective-mass123}}}
\end{figure}

\begin{figure}

\begin{centering}
\includegraphics[scale=0.45]{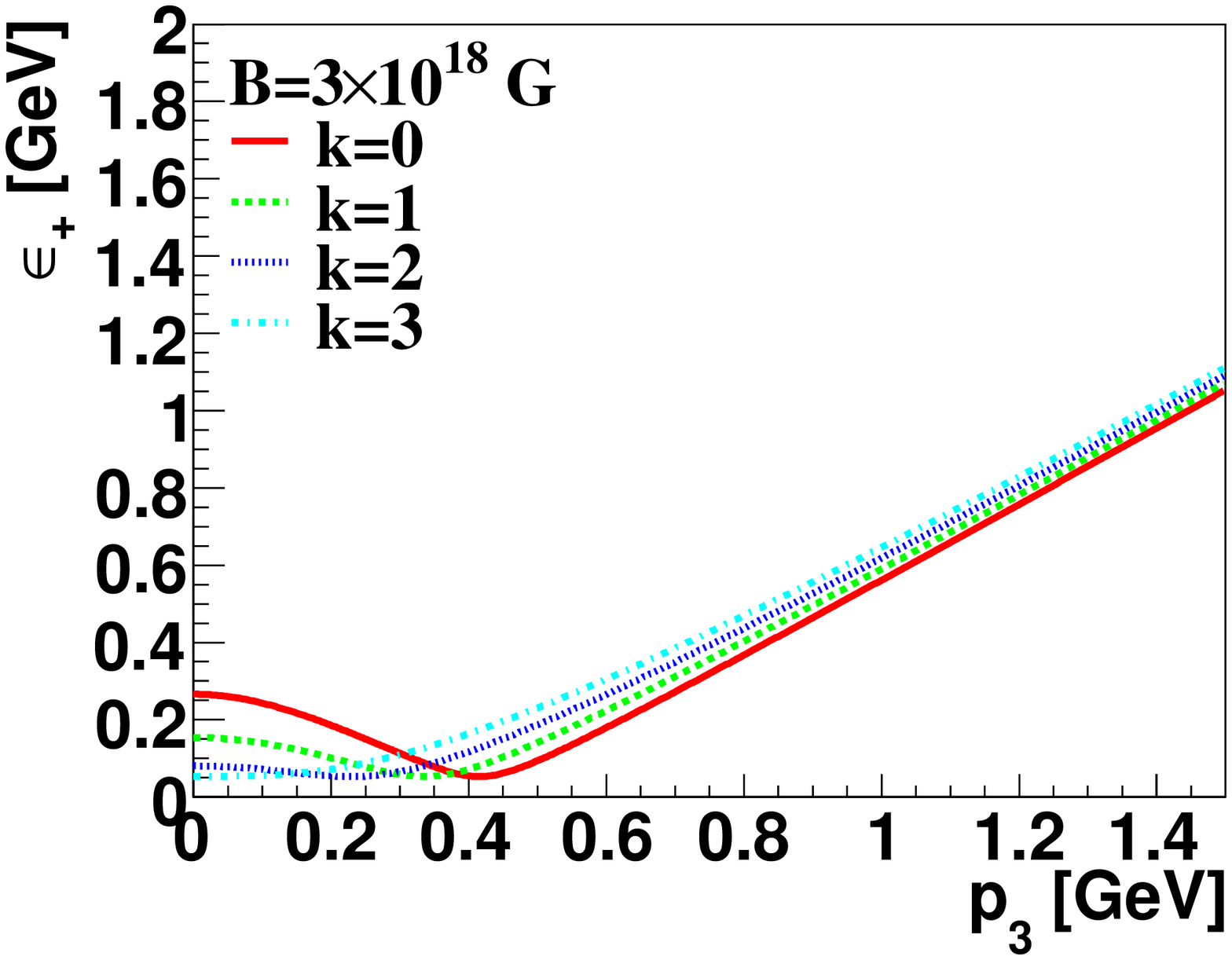}
\par\end{centering}

\begin{centering}
\includegraphics[scale=0.45]{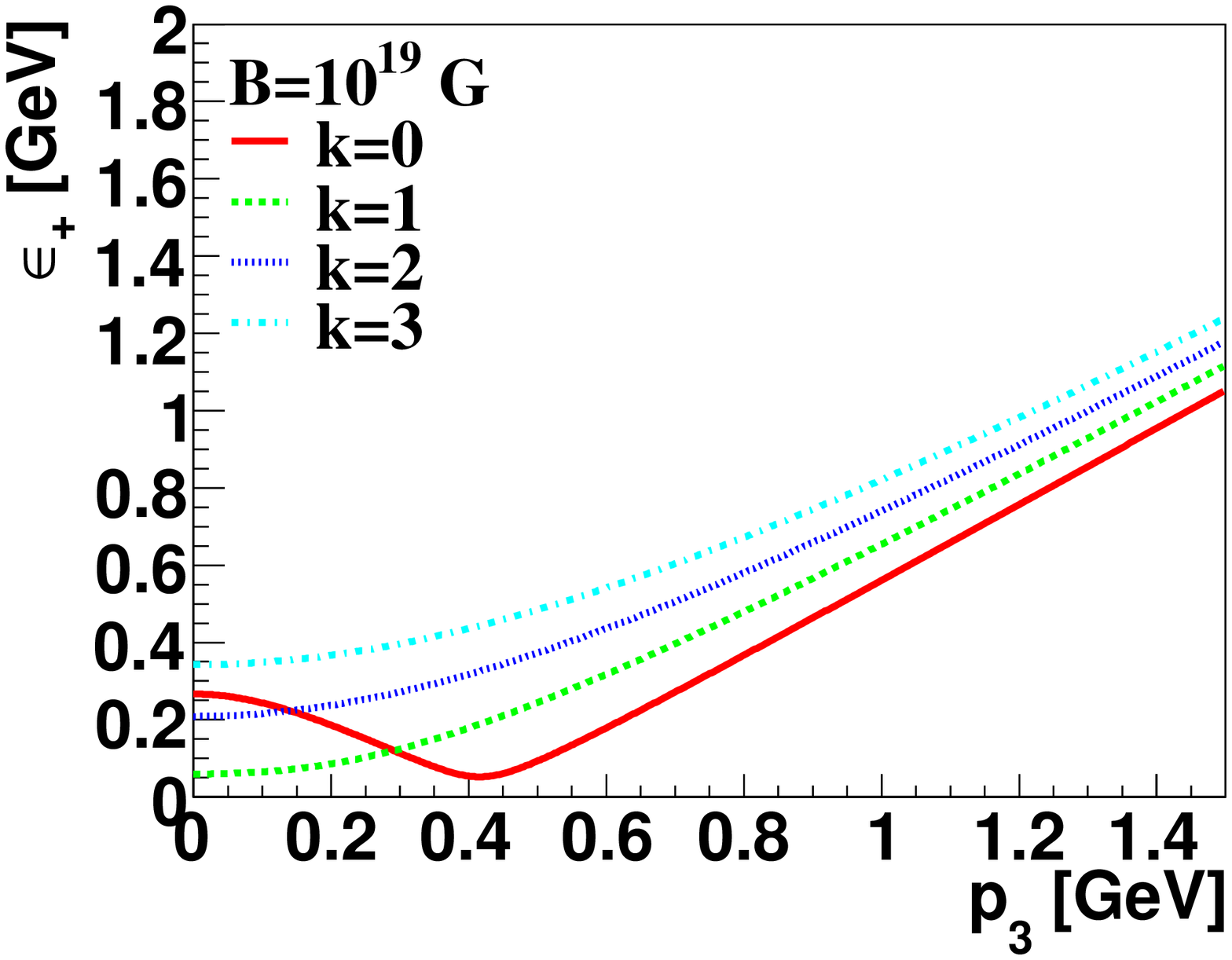}
\par\end{centering}

\begin{centering}
\includegraphics[scale=0.45]{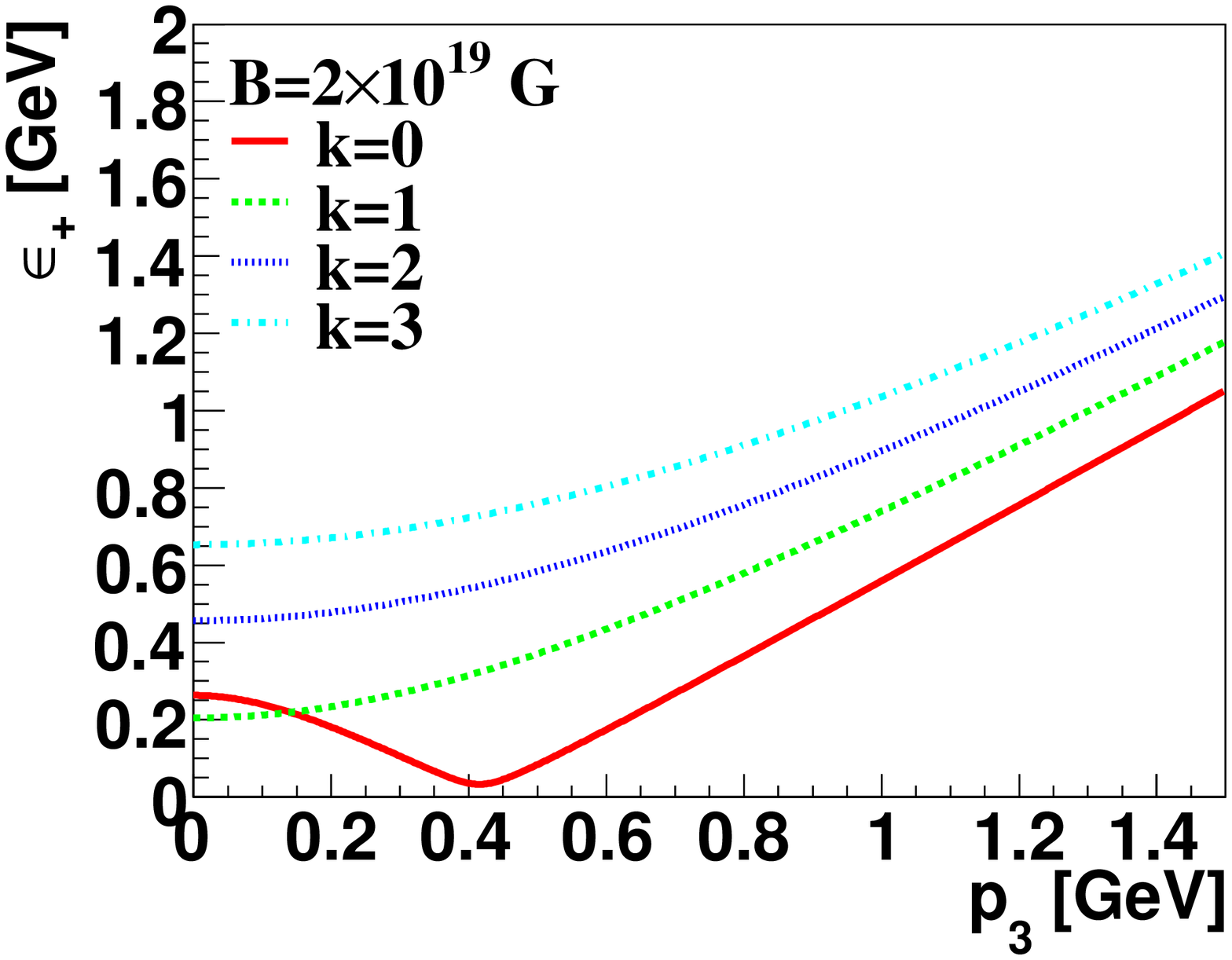}
\par\end{centering}

\caption{{Fermion dispersion relations at three fixed values of magnetic field,
$3\times10^{18}$G, $10^{19}$G and $2\times10^{19}$G. $k=0,1,2,3$
denote the Landau levels. The boson mass $m_{b}$ is set to 1 GeV.
\label{dispersion-relation}}}
\end{figure}

The origin of the crossover to a pure
BCS state at large fields can be understood in terms of the
behavior of the fermion quasiparticle dispersion relations
(\ref{spectrum-f}). To see this, let us introduce the LL-dependent
mass square $M_{k}^{2}\equiv2|qB|k+m^{2}$ in terms of which the
quasiparticle dispersion becomes
\begin{equation}
\epsilon_{+}(k)=\sqrt{(\sqrt{p^{2}_{3}+M_{k}^{2}}-\mu)^{2}+\Delta^{2}}.
\label{particledispersion}\end{equation} Notice that for all the LLs satisfying the condition $\mu>M_{k}$,
the minimum of the dispersion $\epsilon_{+}(k)$ occurs at
$p_{3}=\sqrt{\mu^{2}-M_{k}^{2}}$, with excitation energy given by
the gap $\Delta$, a behavior characteristic of the BCS regime. On
the other hand, for LLs with $\mu<M_{k}$, the minimum of
$\epsilon_{e}(k)$ occurs at $p_{3}=0$, with excitation energy
$\sqrt{(\mu-M_{k})^{2}+\Delta^{2}}$, typical of the BEC regime.
Therefore, the BCS-BEC crossover in the presence of the magnetic
field is controlled by the relative numbers of LLs for which the
sign of the effective chemical potential $\mu_{k}=\mu-M_{k}$ is
either positive (BCS type) or negative (BEC type). In other words,
all the LLs up to certain $k_{BCS}=N$, such that
$M_{N}<\mu<M_{N+1}$, produce fermionlike modes and thus contribute
to the BCS component, while all the LLs with $k>k_{BCS}$ produce
bosonlike modes, hence contributing to the BEC one. When the field
changes, the effective mass $M_k$ changes, as does the total
number of LLs contributing to Eqs. (\ref{numberequa}) and
(\ref{Gap-Eq}) and the density of states of each LL. This leads to
oscillations in the chemical potential $\mu$ that in turn are
reflected in the number of LLs contributing to each regime. At
fields large enough to put all the fermions in the LLL, one has
$M_{k}=m$ and the dispersion reduces to
\begin{equation}
\epsilon_{+}(0)=\sqrt{(\sqrt{p^{2}_{3}+m^{2}}-\mu)^{2}+\Delta^{2}},
\label{LLLdispersion}\end{equation}
thus the system is in the BCS regime, as long as $\mu>m$. Therefore,
a strong enough magnetic field will ultimately favor the crossover
to a BCS regime.

Figure \ref{effective-mass123} shows the consistency of the above interpretation. In this figure we plotted the chemical potential $\mu$ and the effective mass $M_{k}$ of a few LLs as functions
of the magnetic field when the system starts at zero field in the BEC regime (upper panel), or in the BCS one (lower panel). At low fields, the effective masses of all the LLs in the lower panel lie below $\mu$, indicating the predominance of the BCS-type modes (positive $\mu_k$), in agreement with the low-field behavior depicted in Fig. \ref{fractionsversusB2} for the same value of $m_b$. In contrast, the effective masses of most of the LLs in the upper panel either lie over $\mu$ at low fields or cross the $\mu$ line at much smaller field values than in the lower panel case, because the system is throughout this entire small-field range in the BEC side, in agreement with the behavior shown in Fig. \ref{fractionsversusB}. When the field increases the levels start to cross the $\mu$ line, but at the same time the higher LLs start to empty because the density of states of each level increases with the field and hence the lower levels can accommodate more and more fermions. Finally, when the magnetic and the Fermi energies become comparable,  $|qB|\geq \mu^{2}$, all the fermions lie in the LLL with a  BCS-like dispersion because $M_0 < \mu$. Therefore, in the strong field region, all the modes are LLL modes and the two systems lie deep in the BCS regime. This is the basis of the field-induced crossover mechanism.

We call to the reader's attention that, on a closer look, the essence of the
above description of a field-induced relativistic BCS-BEC crossover
is not too different from the essence of the crossover at zero field
previously studied in other systems, relativistic \cite{Sun
etal:2008} and nonrelativistic \cite{Chen etal 2005}. In the
nonrelativistic case \cite{Chen etal 2005}, the crossover can be
induced by the change in the sign of the chemical potential with
increasing coupling that leads to a change in the character of the
system from fermionic $(\mu>0)$ to bosonic $(\mu<0)$. In a similar
fashion, the crossover in  the relativistic system considered in
\cite{Sun etal:2008} is driven by the charge density change and
controlled by the sign of the parameter $\mu-m$ rather than $\mu$
itself \cite{Sun etal:2008}. Notice that what the magnetic field
does in the relativistic case studied in the present paper is to
introduce a new energy scale in the system, $\sqrt{|qB|}$, which
enters in the effective chemical potential $\mu_{k}$. Then the sign
of $\mu_{k}$ and the number of LLs with each sign serve to control
the crossover. The system will be in: a) a BCS state if the majority
of occupied LLs have $\mu_{k}>0$, b) a BEC state if the majority
of LL have $\mu_{k}<0$, or c) in the crossover region if the
numbers of LLs with positive and negative $\mu_{k}$ are comparable.

In Fig. \ref{dispersion-relation} we show the dispersion relations
for quasiparticles $\epsilon _{+}$  for different LLs at three
magnetic fields' values. In the upper panel of
Fig. \ref{dispersion-relation}, the LLs with $k<3$ contribute to the
BCS component, while the one with $k=3$ contributes to the BEC one.
Since the fermion energy splitting between different LLs and the
density of states of each LL are each proportional to $\sqrt{eB}$,
when the field increases not only do the levels become more separated
in energy, as seen from the figure, but also they can accommodate a
larger number of particles. As a consequence, when the field
increases, the number of occupied LLs reduces. This means that the higher levels
shown in the middle and lower panels of Fig.
\ref{dispersion-relation} are likely not contributing already at
those strong fields.

\section{Concluding Remarks}
In this paper we investigated the effect of a magnetic field in the
relativistic BCS-BEC crossover in the context of a model with neutral bosons and
charged fermions minimally coupled to a magnetic field. The simple model used in our calculations resembles some
basic properties of spin-zero color superconducting phases in a
magnetic field like for instance the MCFL phase \cite{MCFL}.

Our results demonstrate that a magnetic field can tune the BCS-BEC
crossover via a novel mechanism according to which the state of the
system at each field is determined by the number of occupied LLs
with either positive or negative effective chemical potential
$\mu_{k}=\mu-M_{k}$. If the majority of the LLs have $\mu_{k}<0$,
the system is in a BEC state, because a majority of BEC-type modes
prevails. On the contrary, if the majority of the LLs have
$\mu_{k}>0$, it is in the BCS regime. At strong enough fields, the
system goes to the BCS regime, because in this case only the LLL,
whose dispersion is always of BCS type, is occupied.

The BCS-BEC crossover has been studied in the literature using two
types of models: single-channel models and two-channel models. In
two-channel models, like the one used in this paper, fermion and
boson degrees of freedom are introduced from the beginning in the
Lagrangian. In single-channel models one starts with a Lagrangian
that only has fermionic degrees of freedom, like in a
Nambu-Jona-Lasinio (NJL) model. Then, the bosonic degrees of freedom
are introduced with the help of a bosonization procedure as the
Hubbard-Stratonovich transformation. A natural continuation of the
present work will be to investigate the magnetic field effect on the
crossover in the context of a single-channel theory. Given that the
dispersions of the charged fermions in the presence of a magnetic
field will be of the same form in a purely fermionic theory in the
presence of a magnetic field, one can still use the sign of the
effective chemical potential $\mu_{k}$ and the relative numbers of
LLs with each sign as valid criteria to control the crossover.

There are different NJL theories that could be used as
single-channel models. One interesting possibility would be to
consider the field effects in a NJL model with both diquark and
chiral condensates that can interact via the axial anomaly, such as the
one considered in \cite{Basler:2010xy}.  We could then explore how
the field-induced crossover mechanism found in our two-channel model
turns out to be in this case, where the mass of the charged fermions
is not necessarily constant, but can itself be affected by the field
in any region where the diquark and chiral condensates coexist.

Apart from the obvious fundamental motivation of understanding the
effects of a magnetic field in the BCS-BEC crossover within a more
realistic model, if the relativistic BCS-BEC theories discussed in
the literature have any relevance for the physics of neutron stars
and the future low-energy, heavy-ion collision experiments, it makes
sense to consider them with a magnetic field, as extremely strong
magnetic fields are expected to be present in these two settings.
Therefore, an imperative next step will be to consider more
realistic models of color superconducting quark matter to explore
all the implications of a magnetic field in the BCS-BEC crossover.

\textbf{Acknowledgments:} The work of VI and EJF has been supported
in part by DOE Nuclear Theory Grant No. DE-SC0002179. QW is supported
in part by the "100 talents" project of the Chinese Academy of Sciences
(CAS) and by the National Natural Science Foundation of China (NSFC)
under Grant No. 10735040.

\end{document}